\documentclass[aps,superscriptaddress,showpacs,floatfix,preprint,showkeys,pre]{revtex4}
\usepackage{times}
\usepackage{xspace}
\usepackage{graphicx,morefloats}
\usepackage{color}
\usepackage{amsmath, amsthm, amssymb}
\usepackage{enumerate}

\begin{document}

\title{The relation between the statistics of open ocean currents
  and the temporal correlations of the wind-stress
}
\author{Golan Bel}
\affiliation{Department of Solar Energy and Environmental Physics, Blaustein Institutes for Desert Research, Ben-Gurion University of the Negev, Sede Boqer Campus 84990, Israel}
\email{bel@bgu.ac.il}
\author{Yosef Ashkenazy}
\affiliation{Department of Solar Energy and Environmental Physics, Blaustein Institutes for Desert Research, Ben-Gurion University of the Negev, Sede Boqer Campus 84990, Israel}
\date{\today}
\begin{abstract}
  We study the statistics of wind-driven open ocean currents. Using
  the Ekman layer model for the integrated currents, we investigate,
  analytically and numerically, the relation between the wind-stress
  distribution and its temporal correlations and the statistics of the
  open ocean currents. We find that temporally long-range correlated
  wind results in currents whose statistics is proportional to the
  wind-stress statistics. On the other hand, short-range correlated
  wind leads to Gaussian distributions of the current components,
  regardless of the stationary distribution of the winds, and
  therefore, to a Rayleigh distribution of the current amplitude, if
  the wind-stress is isotropic. We find that the second moment of the
  current speed exhibits a maximum as a function of the correlation
  time of the wind-stress for a non-zero Coriolis parameter.  The
  results were validated using an oceanic general circulation model.
  
\end{abstract}
\maketitle
\section{Introduction}
\label{intro}
Ocean currents are generated by local and remote forces and factors,
including winds, tides, buoyancy fluxes and various types of
waves. While many studies have investigated the distribution of the wind \cite[][]{Seguro-Lambert-2000:modern,Monahan-2006:probability,Monahan-2010:probability}, 
focusing on its relevance to energy production, the distribution of
ocean currents has received much less attention
\cite[][]{Chu-2008:probability}. Moreover, currently there is no accepted
theory explaining the observed statistics of surface ocean currents.

\begin{figure}
 \centering
 \includegraphics[width=\textwidth]{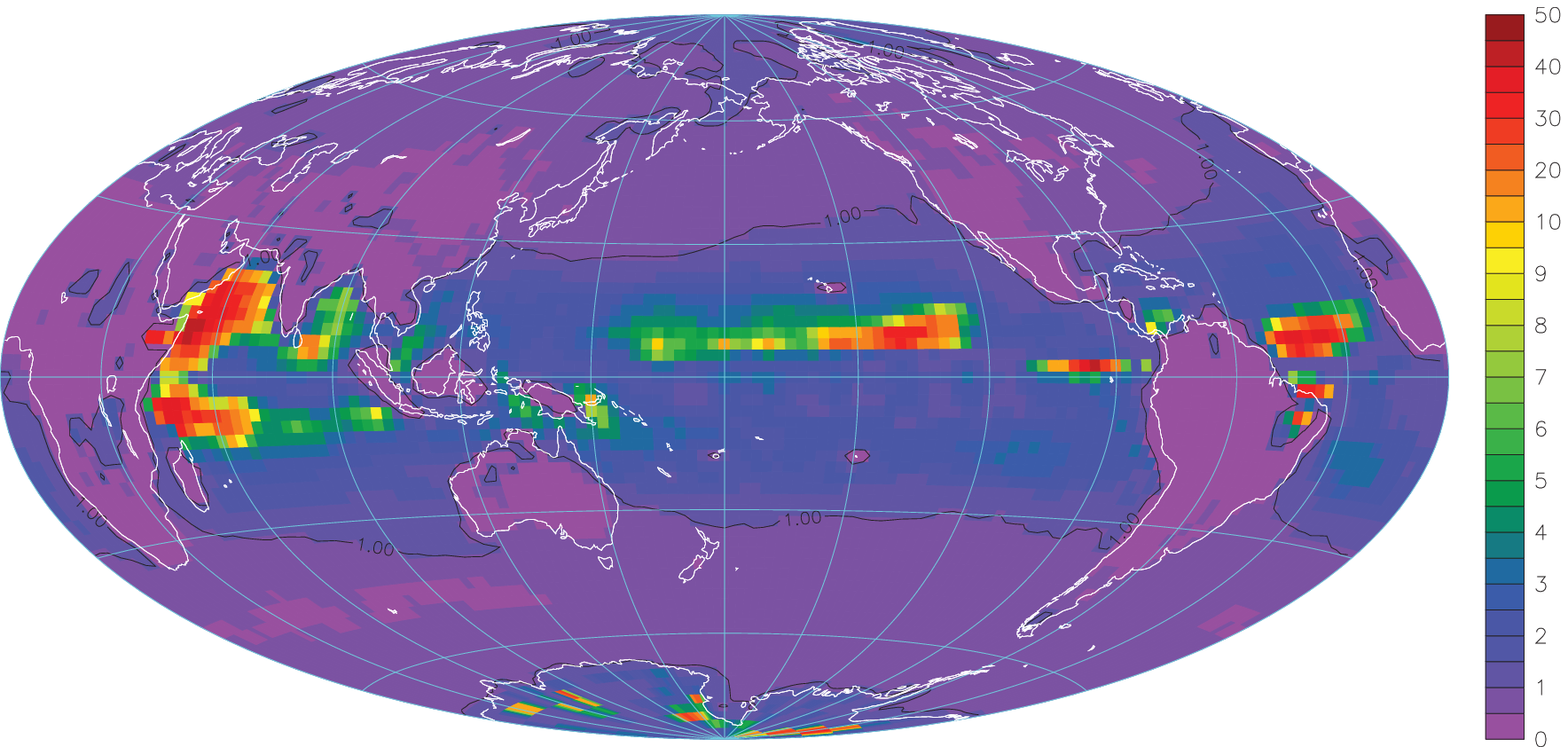}
 \caption{A map of the correlation time (in days) of the wind-stress
   magnitude. We define the correlation time as the time at which the
   normalized auto-correlation function first drops below
   $1/e$. Clearly, there is a large variability in the correlation
   time. One can also notice the remarkably shorter correlation time
   overland compared with that over the ocean (except for the polar regions). The map is based on the
   NCEP/NCAR reanalysis six-hourly wind data for the period of 1993-2010
   \protect\cite[][]{Kalnay-Kanamitsu-Kistler-et-al-1996:ncep}. The spatial resolution of the data is $2.5^\circ\times2.5^\circ$.
  Other definitions of the correlation time yield qualitatively the same results. The
   black contour line indicates a one-day correlation time, and the white line indicates the coast line. 
A grid (light blue
   lines) of $30^\circ\times30^\circ$ was superimposed on the map.}
 \label{fig:1}
\end{figure}

Here, we propose a simple physical theory for the distribution of
wind-driven ocean currents and its relation to the spatially variable
temporal correlations of the wind (see Fig. \ref{fig:1}). We show
that the distribution of wind-driven ocean currents strongly depends
on the temporal correlations of the wind--when the wind exhibits
long-range temporal correlations, the ocean current statistics is
proportional to the wind-stress statistics, while for short-range
correlations of the wind, the different components of the current
vector follow Gaussian distributions. It was previously reported that
the probability density function (PDF) of ocean currents follows the
Weibull distribution
\cite[][]{Gille-Smith-1998:probability,Gille-Smith-2000:velocity,Chu-2008:probability,Chu-2009:statistical,Ashkenazy-Gildor-2011:probability}
(see Section \ref{sec:weibull} for the details of the Weibull distribution);
We argue that this is not necessarily the case even if the wind-stress magnitude
is Weibull distributed.

Oceans play a major and important role in the climate system, and ocean
circulation underlies many climate phenomena, from scales of meters to
thousands of kilometers and from scales of minutes to decades. The gap
in our understanding of ocean current statistics leads to a lack in
understanding the statistics of other related climate
variables. 
Filling in this gap will be useful in many fields: it may
help to predict extreme current events and, thus, may help to securely
design maritime-associated structures. In light of the increasing
efforts to find alternative sources of energy and the idea of using
ocean currents as such a source, knowledge of the currents'
PDF may help to better estimate the energy production and to appropriately
design ocean current turbines that will withstand even extreme current
events. Moreover, knowledge regarding current statistics and, in
particular, its relation to the wind-stress statistics may improve
the parametrization of small-scale processes in state-of-the-art
general circulation models (GCMs).

The study of wind-driven ocean currents goes back more than 100 years,
to the time when Ekman \cite{Ekman-1905:influence} proposed his classical
simple model to explain the effect of the Earth's rotation on upper ocean
currents. His model predicted that the depth-integrated current vector
is perpendicular to the wind vector, a prediction that was largely
proven by observations. Since then, many studies have used Ekman's
model to propose more realistic models for ocean currents, as well as
for surface winds
\cite[][]{Gill-1982:atmosphere,Cushman-Roisin-1994:introduction}.  In
what follows, we use Ekman's model \cite[][]{Ekman-1905:influence} to
study the statistics of wind-driven ocean currents.

This paper is organized as follows. In section \ref{sec:model}, we
describe the Ekman layer model and provide a general (implicit)
solution expressed in terms of the characteristics of the wind-stress
statistics. In section \ref{sec:IC}, we consider two idealized cases
of wind-stress statistics (step-like temporal behavior of the
wind-stress in section \ref{sec:SLws} and exponentially decaying
temporal wind-stress correlation in section \ref{sec:OUws}) and
present analytical expressions for the second moment (and fourth
moment for the first case) of the currents' distribution.  In section
\ref{sec:mitGCM}, we provide a description of the simple numerical model
and of the oceanic GCM (MITgcm) used to validate and extend the
analytical results.
The numerical tests involve the Weibull distribution; hence, a brief
review of the distribution properties and the methods we used to generate correlated and uncorrelated Weibull-distributed time series is provided in \ref{sec:weibull}.  The numerical
results are presented and discussed in section \ref{sec:res}, followed
by a brief summary in section \ref{sec:summary}.

\section{The Ekman model}
\label{sec:model}
In spite of the simplicity of the Ekman model,
it will enable us to start investigating the coupling between the wind-stress
and the ocean currents.  The time, $t$, and depth, $z$,
dependent equations of the Ekman model \cite[][]{Ekman-1905:influence},
describing the dynamics of the zonal (east-west) $U$ and meridional
(south-north) $V$ components of the current vector, are:
\begin{align}\label{UVdyn}
\frac{\partial U}{\partial t}&=fV+\nu\frac{\partial^2U}{\partial z^2} \nonumber  \\
\frac{\partial V}{\partial t}&=-fU+\nu\frac{\partial^2V}{\partial z^2},
\end{align}
where $f=(4\pi/T_{d})\sin(\phi)$ is the local Coriolis parameter
($T_{d}$ is the duration of a day in seconds and $\phi$ is the
latitude), and $\nu$ is the parametrized eddy viscosity coefficient, assumed to be
depth-independent.  In eqs. \eqref{UVdyn}, we consider $U=\tilde
U-U_g$, $V=\tilde V- V_g$ where $U_g$, $V_g$ are the bottom ocean
geostrophic currents determined by the pressure gradient and $\tilde
U$, $\tilde V$ are the actual surface current components. In what
follows, we consider the statistics of $U$, $V$, where the statistics of
$\tilde U$, $\tilde V$ can be obtained by a simple transformation.  The
boundary conditions are chosen such that the current derivative, with
respect to the depth coordinate $z$, is proportional to the integrated
current at the bottom of the layer described by our model and to the
wind-stress vector $(\tau_x,\tau_y)$ at the surface
\cite[][]{Gill-1982:atmosphere,Cushman-Roisin-1994:introduction},
\begin{align}\label{UVbc}
\frac{\partial U}{\partial z}\Big\vert_{z=-h}=\frac{r}{\nu}u(t);\ \ \ \ \ \ \ \ \ \ \ \  \frac{\partial U}{\partial z}\Big\vert_{z=0}=\frac{\tau_x}{\rho_0\nu}; \nonumber \\
\frac{\partial V}{\partial z}\Big\vert_{z=-h}=\frac{r}{\nu}v(t);\ \ \ \ \ \ \ \ \ \ \ \  \frac{\partial V}{\partial z}\Big\vert_{z=0}=\frac{\tau_y}{\rho_0\nu},
\end{align}
where,
\begin{align}\label{defuv}
u\equiv{\displaystyle{\int\limits_{-h}^0U(z)dz}};\ \ \ \ \ \ \ \ \ \ \ \ v\equiv{\displaystyle{\int\limits_{-h}^0V(z)dz}}.
\end{align}
Here, we introduce the following notation: $h$ is the depth of the
upper ocean layer, $r$ is a proportionality constant representing the
Rayleigh friction
\cite[][]{Airy-1845:tides,Pollard-Millard-1970:comparison,Kundu-1976:analysis,Kase-Olbers-1979:wind,Gill-1982:atmosphere},
$(\tau_x,\tau_y)$ are the wind-stress components, and $\rho_0$ is the
ocean water density (hereafter assumed to be constant,
$\rho_0=1028kg/m^3$). The value used for $r$ in the numerical
calculations is based on the empirical estimate outlined in
\cite{Simons-1980:circulation,Gill-1982:atmosphere}.  

By integrating eqs.  \eqref{UVdyn} over a sufficiently deep layer
(i.e., $h\gg\sqrt{2\nu/|f|}$), we obtain the equations describing the
depth integrated currents and their coupling to the wind-stress,
\begin{align}\label{uvdyn}
\frac{\partial u}{\partial t}&=fv-ru+\frac{\tau_x}{\rho_0}, \nonumber  \\
\frac{\partial v}{\partial t}&=-fu-rv+\frac{\tau_y}{\rho_0}.
\end{align}
In the derivation above, we did not explicitly consider the pressure gradient.
Explicit inclusion of the pressure gradient would result in constant
geostrophic currents, and equations \eqref{uvdyn} describe the dynamics
of the deviation from the geostrophic currents. 

To allow a simpler treatment of these equations, we define $w\equiv
u+iv$.  It is easy to show that $w$ obeys the following equation:
\begin{align}\label{wdyn}
\frac{\partial w}{\partial t}&=-ifw-rw+\frac{\tau}{\rho_0},
\end{align}
where $\tau\equiv \tau_x+i\tau_y$. Eq. \ref{wdyn} is a complex Langevin equation in which the complex noise is not necessarily Gaussian \cite{vanKampen-1981:stochastic}.  The formal solution of equation
\eqref{wdyn} is
\begin{align}\label{wsol}
  w(t)&=w(0)e^{-(if+r)t}+\frac{1}{\rho_0}{\displaystyle{\int\limits_0^t\tau(t')e^{-(if+r)(t-t')}dt'}}.
\end{align}
This formal solution demonstrates that the currents depend on the
history of the wind-stress, and therefore, the distribution of the
currents depends, not only on the  wind-stress distribution, but on all
multi-time moments of the wind-stress.

However, the two extreme limits are quite intuitive.  When the
correlation time of the wind-stress, $T$, is long (i.e., $T\gg 1/r$
and $T\gg 1/f$), one expects that the currents will be proportional to
the wind-stress since the ocean has enough time to adjust to the wind
and to almost reach a steady state. In terms of eq. \eqref{wsol}, in
this limit, $\tau(t')$ can be approximated by $\tau(t)$ since its
correlation time is longer than the period over which the exponential
kernel is non-zero. In the other limit, when the correlation time of
the wind-stress is very short (i.e., $T\ll 1/r$ and $T\ll 1/|f|$), namely,
the wind-stress is frequently changing in a random way, the ocean is
not able to gain any current magnitude. In this case, the central
limit theorem \cite[][]{vanKampen-1981:stochastic} implies that each component of the
current vector is Gaussian distributed.

It is useful to write the formal expression for the square of the
current’s amplitude in terms of the wind-stress temporal correlation
function.  Taking the square of eq. \eqref{wsol} and averaging over
all realizations (with the same statistical properties) of the wind
stress, one obtains:
\begin{align}\label{w2}
&\langle |w(t)|^2 \rangle =|w(0)|^2 e^{-2rt}\\
& +\frac{w(0)^*}{\rho_0} e^{-(r-if)t}{\displaystyle{\int\limits_0^t\langle\tau(t')\rangle e^{-(if+r)(t-t')}dt'}} \nonumber \\
&+\frac{w(0)}{\rho_0} e^{-(r+if)t}{\displaystyle{\int\limits_0^t\langle\tau(t')^*\rangle e^{-(r-if)(t-t')}dt'}} \nonumber \\
&+\frac{1}{\rho_0^2}{\displaystyle{\int\limits_0^t\int\limits_0^t \langle \tau(t')\tau(t'')^*\rangle e^{-if(t''-t')}e^{-r(2t-t'-t'')}dt'dt''}}. \nonumber
 \end{align}
The variance of the current’s amplitude may be written as
\begin{align}\label{wvar}
	&\langle |w(t)|^2 \rangle-|\langle w(t)\rangle|^2=\\
	&\frac{1}{\rho_0^2}{\displaystyle{\int\limits_0^t\int\limits_0^t C(t',t'') e^{-if(t''-t')}e^{-r(2t-t'-t'')}dt'dt''}},\nonumber
\end{align}
where we defined the temporal correlation function of the wind-stress as
\begin{align}\label{CF}
	C(t',t'')=\langle \tau(t')\tau(t'')^*\rangle-\langle \tau(t')\rangle\langle \tau(t'')^*\rangle.
\end{align}

To allow analytical treatment, we proceed by considering two special
idealized cases in which the formal solution takes a closed analytical
form. For simplicity, we only consider the case of statistically
isotropic wind-stress \cite[the direction of the wind-stress is
uniformly distributed,][]{Monahan-2007:empirical}, namely a case in which the
 wind-stress components are independent, identically distributed
 variables with identical temporal autocorrelation functions. 
The latter assumption is not always valid \cite[see for
example,][]{Monahan-2012:temporal}; however, the generalization of our
results to the case of non-isotropic wind stress is straightforward.

\section{Idealized cases}
\label{sec:IC}
\subsection{Step-like wind-stress}
\label{sec:SLws}
A simple way to model the temporal
correlations of the wind is by assuming that the wind vector randomly changes
every time period $T$ while remaining constant
between the ``jumps.'' By integrating the solution of $w$
(eq. \eqref{wsol}) and taking into account the fact that the
distribution of $w$ at the initial time and final time should be
identical, one can obtain the expressions for the second and fourth
moments (ensemble average over many realizations of the stochastic
wind-stress) of the current amplitude. To better relate our results
to the outcome of data analysis, we also need to consider the time
average since the records are independent of the constant wind-stress period.
The double averaged second moment of the current’s amplitude is defined as
\begin{align}
 \overline{\left\langle\left|w\right|^2\right\rangle}\equiv\left\langle\frac{1}{T}{\displaystyle{\int\limits_0^T|w(t)|^2dt}}\right\rangle,
\end{align}
and the angular brackets represent the average over different realizations of the wind-stress with the same statistical properties.
Using eq. \eqref{w2} and the fact that in this idealized case, the wind stress is constant over the period of the step ($T$), we obtain for
the double averaged second moment of the current’s amplitude (see the supplementary material for the details of the derivation): 
\begin{equation}\label{w2mT}
  \overline{\left\langle\left|w\right|^2\right\rangle}=\frac{\left\langle\left|\tau\right|^2\right\rangle\left(1+\frac{1-A_1}{rT}-2\frac{r\left(1-A_1\right)+fA_2}{T\left(f^2+r^2\right)}\right)}{\left(f^2+r^2\right)\rho_0^2},    
\end{equation}
where we have introduced the notations $A_1\equiv\exp(-r T)\cos(fT)$,
and $A_2 \equiv \exp(-rT)\sin(fT)$.

The two extreme limits discussed in section \ref{sec:model} can be
easily realized and understood. When the correlation time $T$ is very
long compared with $1/r$, the second moment of the current is
proportional to the second moment of the wind-stress, namely, $\langle
\left|w\right|^2\rangle =
\langle\left|\tau\right|^2\rangle/[(f^2+r^2)\rho_o^2]$. This is
exactly what one would expect. The long duration of constant
wind-stress allows the system to fully respond and adjust to the
driving force, and hence, the moments of the currents are proportional
to the moments of the wind-stress.  The coefficient of proportionality
is given by the solution of eqs. \eqref{uvdyn} with constant
wind-stress.  The other limit is when $rT,fT\ll 1$ and, in this case,
$\langle \left|w\right|^2\rangle\sim\langle\left|\tau\right|^2\rangle
T/(2r\rho_0^2)$.

In this limit, the second moment of the current amplitude is very small and approaches zero as the correlation time
approaches zero. This result is what one
would expect for a rapidly changing wind that cannot drive significant currents.
Using eq. \eqref{wsol} and the fact that wind-stress is constant over the step period ($T$), we calculate the double averaged (over the step period and the ensemble of wind-stress realizations) fourth moment (see the supplementary material for the details of the derivation),  
\begin{eqnarray}
\overline{\left\langle \left|w\right|^{4}\right\rangle }&=&\frac{\langle\left|\tau\right|^{2}\rangle^{2}g_2\left(T\right)+\langle\left|\tau\right|^{4}\rangle g_4\left(T\right)}{\left( f^{2}+r^{2}\right)^{2}\rho _{0}^{4}}, \\ 
g_2\left(T\right)&=&\frac{4\left(D+1-2A_{1}\right) }{1-D}\times \nonumber \\
&&\left( \frac{3-D-2A_{1}D}{4rT}-2\frac{3r\left(1-DA_{1}\right) +fDA_{2}}{\left( f^{2}+9r^{2}\right) T}\right), \nonumber \\
g_4\left(T\right)&=&1+\frac{5-3D+2A_1(A_1-1-D)}{2rT} \nonumber \\
&-&4\frac{3r\left( 1-DA_{1}\right)+fDA_{2}}{\left(f^{2}+9r^{2}\right) T} \nonumber \\
&+&\frac{r(1-A_{1}^{2})+rA_{2}^{2}+2fA_{1}A_{2}}{(f^2+r^2)T} \nonumber \\
&-&4\frac{r(1-A_1)+fA_2}{(f^2+r^2)T},\nonumber
\end{eqnarray}
where $D\equiv\exp(-2rT)$.
Here, we again use the fact that the integrals of the time averaging
are easily carried due to the fact that the wind-stress is constant during
the averaged period. 

While at the limit $T\to 0$, both the second and the fourth moments
vanish, the ratio between the fourth moment and the square of the
second moment remains finite and is equal to $2$.  This corresponds to
the Rayleigh distribution which originates in the facts that the
components of the currents are independent and each one has a Gaussian
distribution with a zero mean and the same variance. 

One can easily understand the origin of the Gaussian distribution by
considering the central limit theorem, which can be applied in this
limit. Each of the current components is the integral of many
independent, identically distributed random variables (the wind-stress
at different times).  Based on the second and fourth moments in the
limit of $T\to 0$, we conjecture that the overall PDF of the ocean
current speed, in this case, is given by the Rayleigh distribution. 

In \cite{Gonella-1971:local}, the case of constant wind-stress over
some duration was investigated using the depth-dependent Ekman
model. The existence of a wind-stress duration for which the current
amplitude is maximal was found. The results of this section generalize
these results by considering the randomness of the wind-stress and its
continuity. In addition, we consider here the empirical Rayleigh
friction.  

\begin{figure}
 \centering
 \includegraphics[width=\textwidth]{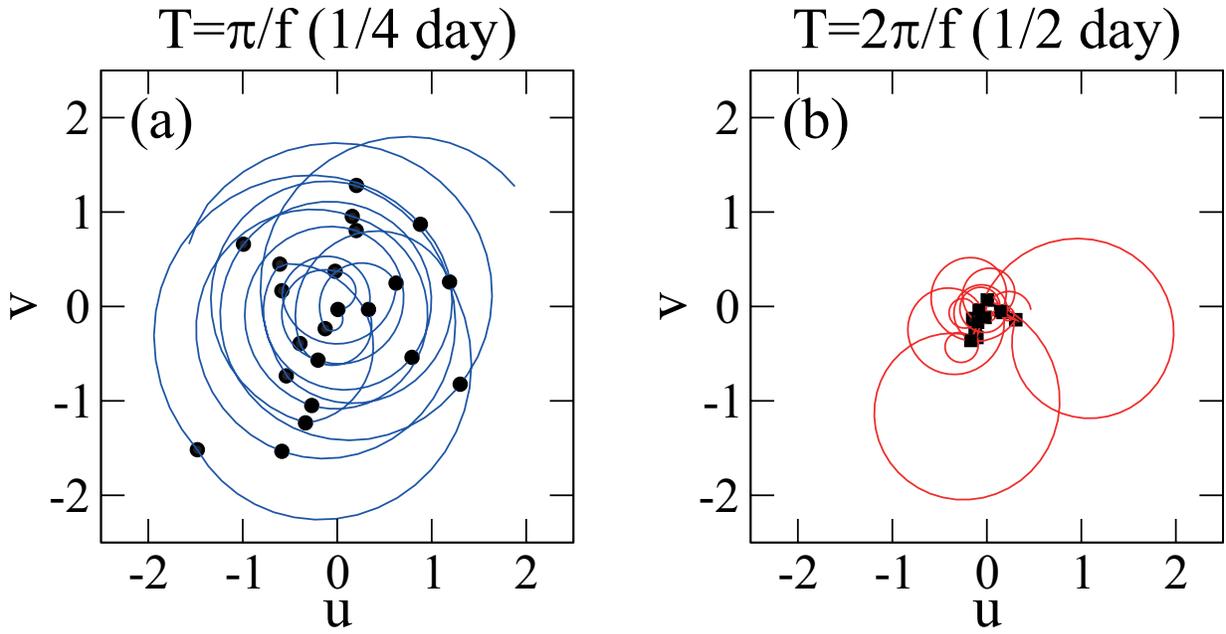}
 \caption{Typical trajectories of the current velocity components for
   the step-like wind stress. In the left panel, (a) the constant
   wind-stress duration, $T=\pi /f$, corresponds to changes in the
   wind-stress after the velocity vector completes $\sim 1/2$
   circle. On the right panel, (b), $T=2\pi /f$, that is, the
   wind-stress changes after a full circle of the velocity field. In
   both panels, the Coriolis parameter, $f$, was set to its value at
   the pole. The symbols indicate the times at which the wind-stress
   was changed.
 }
 \label{fig:2rs}
 \end{figure}

\begin{figure}
 \centering
 \includegraphics[width=\textwidth]{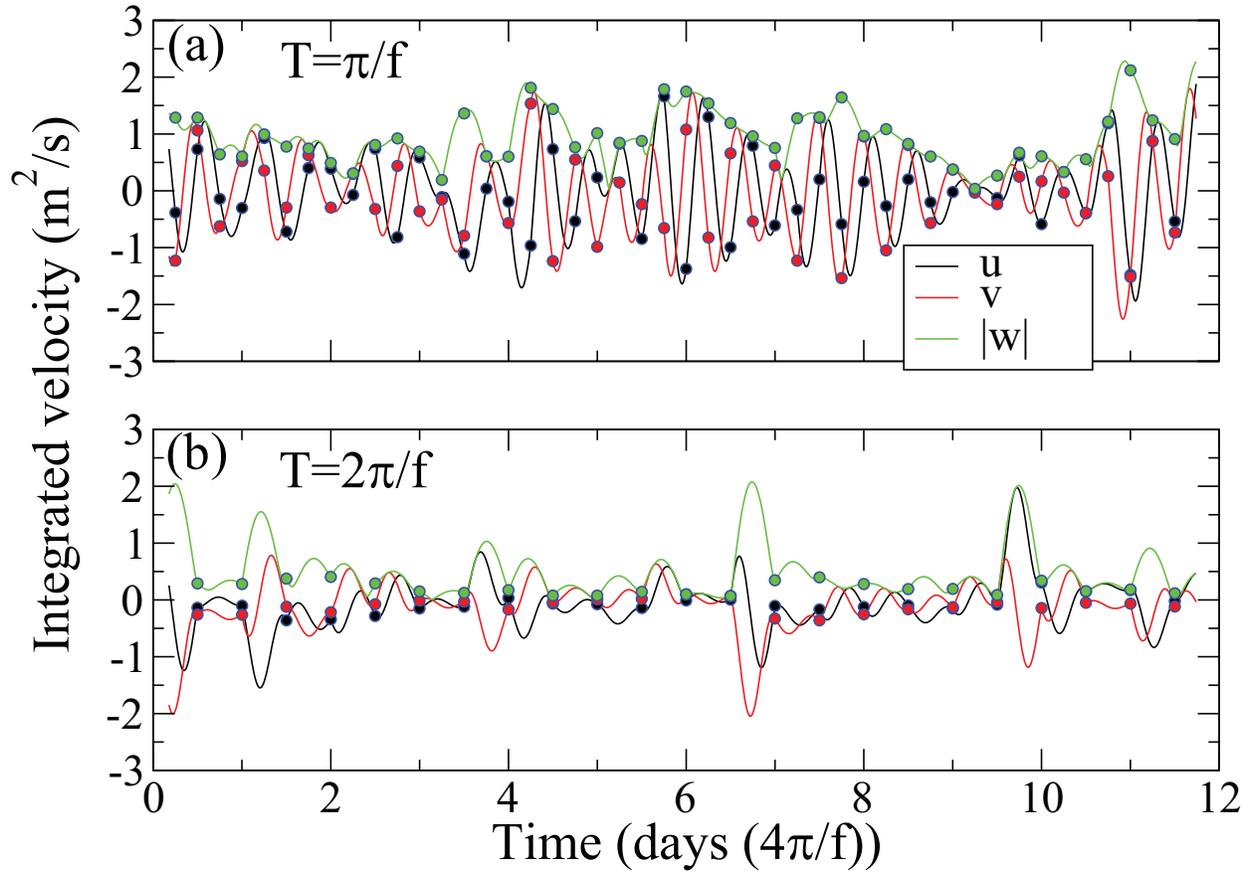}
 \caption{Typical time series of the current velocity components and
   amplitude. In the top panel, (a), the constant wind-stress
   duration, $T=\pi /f$. In the bottom panel, (b), $T=2\pi /f$. In
   both panels, the Coriolis parameter, $f$, was set to its value at
   the pole. The symbols indicate the times at which the wind-stress
   was changed.}
 \label{fig:3rs}
 \end{figure}

In order to better understand the existence of constant wind-stress
duration, $T$, for which the average current amplitude is maximal, we
present, in Fig. \ref{fig:2rs}, two typical trajectories of the current
velocity components $u,v$.  The dynamics of the velocity field may be
understood as follows.  When the friction, $r$, is smaller than the
Coriolis frequency, the velocity field undergoes circular motion and
returns, with frequency $f$, to a point in the velocity space which is
very close to the initial point.  The initial current is irrelevant
since even a small amount of friction ensures that the initial point is close to
the origin at the long-time limit.  The above dynamics is valid for
any value of the temporal wind-stress. Therefore, if the wind-stress
changes with a frequency equal to the Coriolis frequency, the current
amplitude cannot be driven far from the origin, and its average remains
small (this explains the observed minimum for $T\approx 2\pi/ f$). On the
other hand, when the wind-stress changes with a frequency equal to
double the Coriolis frequency, the current velocity reaches the largest
amplitude when the wind-stress changes and the new circle starts
around a different initial point (possibly far from the
origin). Therefore, in the latter case, the average current speed is
maximal.  To better illustrate this picture, we present, in
Fig. \ref{fig:3rs}, typical time series of the current velocity
components and amplitude. The dots represent the times when the
wind-stress changed.  It can be seen that for $T=\pi /f$, the typical
velocity amplitude when the wind-stress changes (the dots in
Fig. \ref{fig:3rs}), is significantly larger than the one for $T=2\pi/
f$.

\begin{figure}
 \centering
 \includegraphics[width=\textwidth]{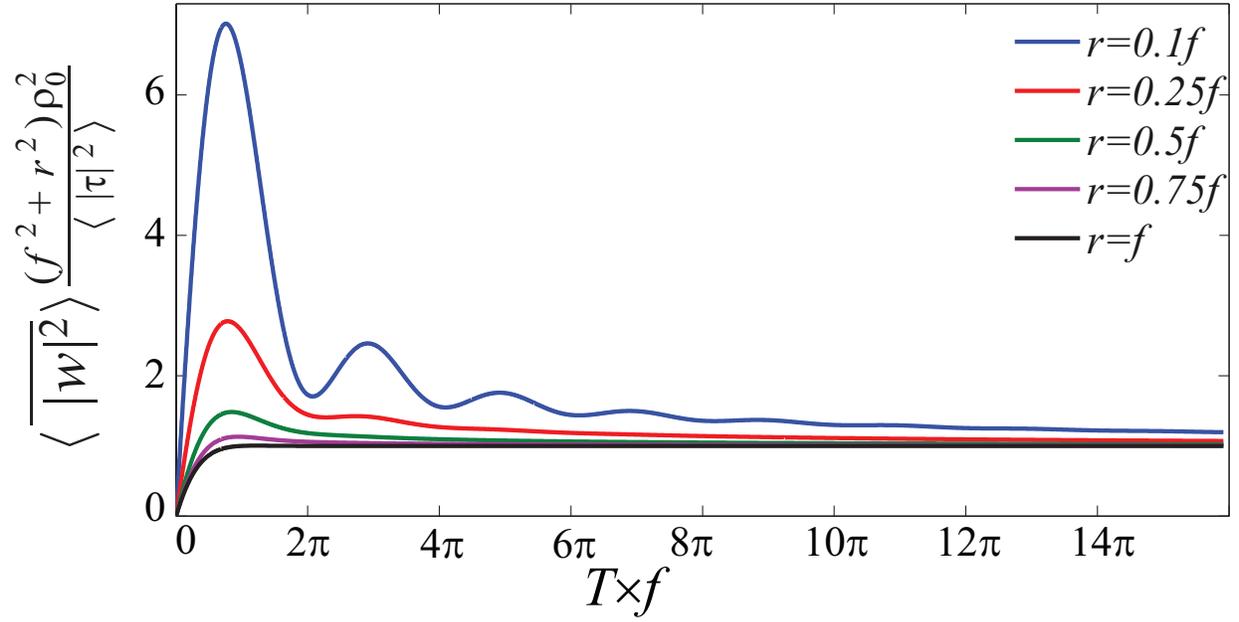}
 \caption{The second moment of the current amplitude
   (eq. \eqref{w2mT}) versus the constant wind-stress duration, $T$.
   The different lines correspond to different values of the Rayleigh
   friction, $r$.}
 \label{fig:4rs}
 \end{figure}

The intuitive explanation above relies on the fact that $r\ll f$. To
elucidate the role of the friction, we present, in Fig. \ref{fig:4rs},
the analytically calculated second moment (eq. \eqref{w2mT}) of the
current versus the constant wind-stress duration for different values
of $r$. When the viscosity is large, there is no maximum but rather a
monotonic increase of the average current amplitude as $T$ increases.  

 \subsection{Exponentially decaying temporal correlations of the wind-stress}
\label{sec:OUws}
The second case we consider is the Ornstein-Uhlenbeck
wind-stress. This process results in a Gaussian wind-stress with an
exponentially decaying temporal correlation function. It is important
to note that, due to the Gaussian nature of the force, the first and the second moments of the wind-stress, together with its two-point correlation function, provide all the information on the driving force. The
correlation function of the wind-stress components is given by
\begin{align}\label{expcf}
 \langle \tau_{i}(t)\tau_{j}(t') \rangle=\delta_{ij}\langle\tau_i^2\rangle\exp(-\gamma_i|t-t'|).
\end{align}
The long-time limits of the current components' averages are:
\begin{align}
 \langle u \rangle\sim\frac{r\langle\tau_x\rangle+f\langle\tau_y\rangle}{\rho_0\left(f^2+r^2\right)};\ \ \ \
\langle v \rangle\sim\frac{r\langle\tau_y\rangle-f\langle\tau_x\rangle}{\rho_0\left(f^2+r^2\right)}.
\end{align}
 The long-time limit of the current amplitude variance,
 $s^2\equiv\langle |w|^2\rangle-\langle u\rangle^2-\langle v
 \rangle^2$, is given by
\begin{align}\label{2mexp}
 s^2\sim{\displaystyle\sum\limits_{i=x,y}}\frac{\langle\tau_i^2\rangle\left(r+\gamma_i\right)}{\rho_0^2r\left(f^2+\left(r+\gamma_i\right)^2\right)}.
\end{align}
Under the assumption of isotropic wind-stress, the variance may be
expressed as
$s^2\sim\langle|\tau|^2\rangle\left(r+\gamma\right)/\left(\rho_0^2r\left(f^2+\left(r+\gamma\right)^2\right)\right)$.
Note that in deriving the expression above, we assumed that the
autocorrelation function of the wind-stress is isotropic.  In general,
this is not the case \cite[][]{Monahan-2012:temporal}; however, for
simplicity, we present here the results for this special case, and the
generalization to the more realistic case is trivial.  Due to the
Gaussian nature of the wind-stress and the linearity of the model, the
current components have Gaussian distributions that are fully
characterized by the mean and variance. Note that the result of
eq. \eqref{2mexp} holds for any distribution of the wind-stress
components, as long as the two-point correlation function is given by
eq. \eqref{expcf}. Equation \eqref{2mexp} is easily derived by
substituting the two-point correlation function (eq. \eqref{expcf}) in
eq. \eqref{wvar}.  Similar to the first idealized case discussed above, it is clear that the second moment vanishes at the limit of a
short wind-stress correlation time, ($\gamma \gg r$ and $\gamma \gg
f$).

\section{Details of the numerical models}
\label{sec:mitGCM}
We performed two types of numerical simulations to validate the
analytical derivations presented above. The first type was a simple
integration of eqs. (\ref{uvdyn})--these numerical solutions are
  provided, both to validate the analytical results and to present the
  solutions of cases not covered by the analytical solutions.
In the second type of simulation, we used a state-of-the-art oceanic GCM, the
MITgcm \cite[][]{MITgcm-manual-2010:mitgcm1}, to test the
applicability of our analytical derivations when the spatial
variability of the bottom topography and the nonlinearity of the ocean
dynamics \cite[][]{Gill-1982:atmosphere} are taken into account. 
\subsection{MITgcm setup and details}
\label{MITgcmdet}
\begin{figure}
 \centering
 \includegraphics[width=\textwidth]{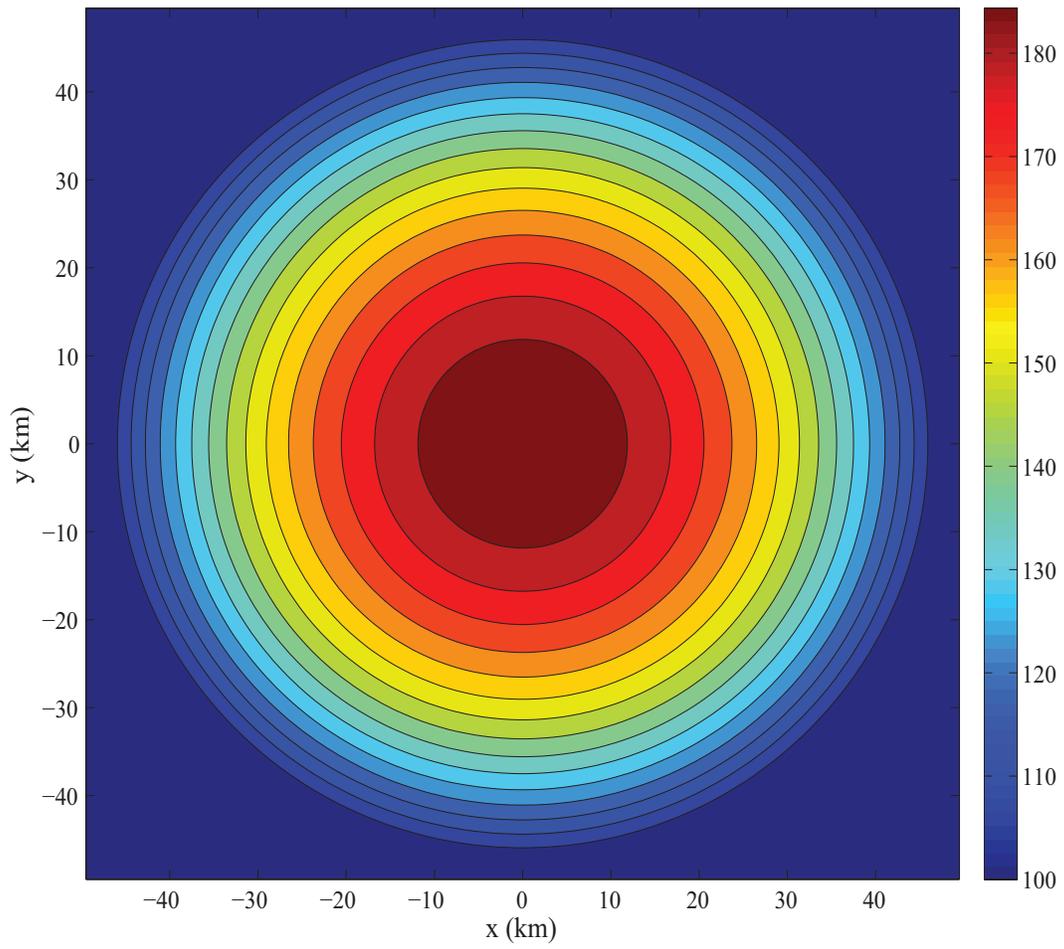}
 \caption{Contour plot of the bathymetry used in the MITgcm simulations.}
 \label{fig:5rs}
 \end{figure}

The MITgcm solves the primitive equations
  \cite[][]{MITgcm-manual-2010:mitgcm1} and is implemented here using
  Cartesian coordinates with a lateral resolution of 1 km (or 10 km)
  with a $50\times50$ grid points. We consider open (periodic) and
  closed physical boundaries of the domain (the results of the closed boundary setup are presented only in the last figure).  There is one vertical
  level that expresses a 2D parabolic basin with a maximum depth of $90 m$, as shown in
  Fig. \ref{fig:5rs}. The basin is situated in a plateau of $100 m$ depth.
 The use of the partial cell option of the MITgcm
  enables us to handle depth variability, even with a single vertical
  level. The integration time step is $10 s$, and the overall
  integration time is two years. Water temperature and salinity are
  kept constant. The horizontal viscosity is $1 m^2/s$, and the vertical
  one is $1\times 10^{-4} m^2/s$. We use the linear bottom drag option
  of the MITgcm, and the effective drag coefficient is depth dependent
  where the mean bottom drag coefficient is $1\times 10^{-5}
  s^{-1}$. We also use the implicit free surface scheme of the MITgcm.

  There are different ways to introduce heterogeneity to the water
  flow, for example, by forcing the water surface with spatially
  variable wind stress. Changes in density due to changes in
  temperature and salinity, as a result of spatially and temporally
  variable surface heat and freshwater fluxes, can also enrich water
  dynamics. The main goal of the MITgcm numerical experiments
  presented here is to validate the analytical results presented above
  when advection and horizontal viscosity are taken into account. To
  enrich water dynamics, we chose to vary the water depth when
  including (excluding) the boundaries of the domain--there is no
  particular reason for choosing this way over another. We find the
  depth variability simpler, from a numerical point of view, but still
  rich enough to allow us to study the boundary effects on the water
  dynamics and to compare the numerical results with the derived
  analytical expressions.

\subsection{The Weibull distribution}
\label{sec:weibull}
It is widely accepted that the PDF of ocean currents follows the
Weibull distribution
\cite[][]{Gille-Smith-1998:probability,Gille-Smith-2000:velocity,Chu-2008:probability,Chu-2009:statistical,Ashkenazy-Gildor-2011:probability};
therefore, we use this distribution in our numerical tests.  For
consistency, we provide here a brief review of its properties and the
methods we used to implement temporally correlated and uncorrelated
time series.  The Weibull PDF is defined for positive values of the
variable, $x>0$, and is characterized by two parameters, $k$ and
$\lambda$:
\begin{equation}\label{weibull}
W_{k,\lambda}(x)=\frac{k}{\lambda} \left(
  \frac{x}{\lambda} \right) ^{k-1} \exp( -\left( x/\lambda \right) ^{k}).
\end{equation}
where $\lambda$ and $k$ are both positive.  $\lambda$ and $k$ are usually
referred to as the scale and shape parameters of the distribution. The
cumulative Weibull distribution function is given by:
\begin{equation}
  F_{k,\lambda}(x) = 1-e^{-(x/\lambda)^{k}}.
\end{equation}
It is possible to express the moments of the Weibull distribution, $\langle x^m \rangle$,
using $k$ and $\lambda$:
\begin{equation}
  \langle x^m \rangle = \lambda^m \Gamma\left(1+\frac{m}{k}\right),
\end{equation}
where $\Gamma$ is the Gamma function. Thus, it is sufficient to
calculate the first and second moments of a time series and, from them,
to find the $k$ and $\lambda$ that characterize the Weibull
distribution (assuming that we have a priori knowledge that the series
is Weibull distributed).  The $k$ parameter can be calculated using
any two different moments ($n$, $m$) of the time series by solving
(numerically) a transcendental equation $\langle x^n
\rangle^{m/n}/\langle x^m\rangle=\left(\Gamma(1+n/k)\right)^{m/n}/\Gamma(1+m/k)$,
and $\lambda$ can be estimated by using the first moment (the mean)
and the estimated $k$.  It is also possible to calculate the $k$
parameter from the slope of the hazard function of the Weibull
distribution $W(x)/(1-F(x))=(k/\lambda)(x/\lambda)^{k-1}$ when a
log-log plot is used. In this manuscript, we use the second and the
fourth moments to derive the values of $k$ and $\lambda$.  Previous
studies used other approximations to estimate $k$ and $\lambda$
\cite[][]{Monahan-2006:probability,Chu-2008:probability}.

A special case of the Weibull distribution, called the Rayleigh
distribution, is obtained when $k=2$. It can be associated with the
distribution of the magnitude of a vector whose components are two
independent, Gaussian-distributed, random variables; i.e., if $x$ and
$y$ are Gaussian-distributed independent random variables, then the
distribution of $s=\sqrt{x^2+y^2}$ is a Rayleigh distribution (Weibull
distribution with $k=2$).

It is fairly easy to generate an uncorrelated Weibull-distributed time
series (using a simple transformation rule). However, below we use time series that are both Weibull
 distributed and temporally correlated.  Such
time series are generated as follows: 
\begin{enumerate}[(i)]
\item Generate uncorrelated, Gaussian-distributed, time series.
\item Introduce temporal correlations by applying Fourier transform to
  the time series from (i), multiply the obtained power spectrum by
  the power spectrum corresponding to the desired correlations (in our case,
  exponentially decaying temporal correlations) and apply an inverse
  Fourier transform.
\item Generate uncorrelated Weibull-distributed time series.
\item Rank order the time series from step (iii) according to the time
  series of step (ii). 
\end{enumerate}
The resulting time series is both temporally correlated and Weibull
distributed. It is important to note that deviations from the desired Weibull parameters and the correlation time may occur due to the finite size of the time series. In order to minimize these deviations, one must ensure that the time series are much longer than the relevant correlation time. Elaborated discussion of this and similar methods may be found in \cite{SchreiberAndSchmidt:1996,SchreiberAndSchmidt:2000,KantzAndSchreiber:2004}.
In deriving the results presented in the following section, we have used the algorithm mentioned above to generate the Weibull-distributed and temporally-correlated wind stress.
In addition, the parameters characterizing the Weibull distribution of the currents ($k_{\rm current}$ and $\lambda_{\rm current}$) were derived using the second and fourth moments and the relations mentioned in this subsection. 
\section{Results}
\label{sec:res}
The analytical results presented above for the idealized cases highlight
the important role played by the temporal correlations of the
wind-stress in determining the statistics of surface ocean currents.
The behaviors at the limits of long-range temporal correlations
($rT\gg 1$) and short-range temporal correlations ($T\to0$) are
intuitive, once derived.  The existence of an optimal correlation
time, at which the average current amplitude is maximal, is less
trivial. A similar maximum was reported by \cite{Gonella-1971:local}, as
described above.  \cite{McWilliams-Huckle-2005:ekman} (right panel of
Fig. 11 of their paper) studied the Ekman layer rectification using
the $K$-profile-parameterization and observed a maximum in the
depth-averaged variance of the current speed as a function of the
Markov memory time. In what follows, we present the results of the
numerical tests.

\begin{figure}
 \centering
 \includegraphics[width=\textwidth]{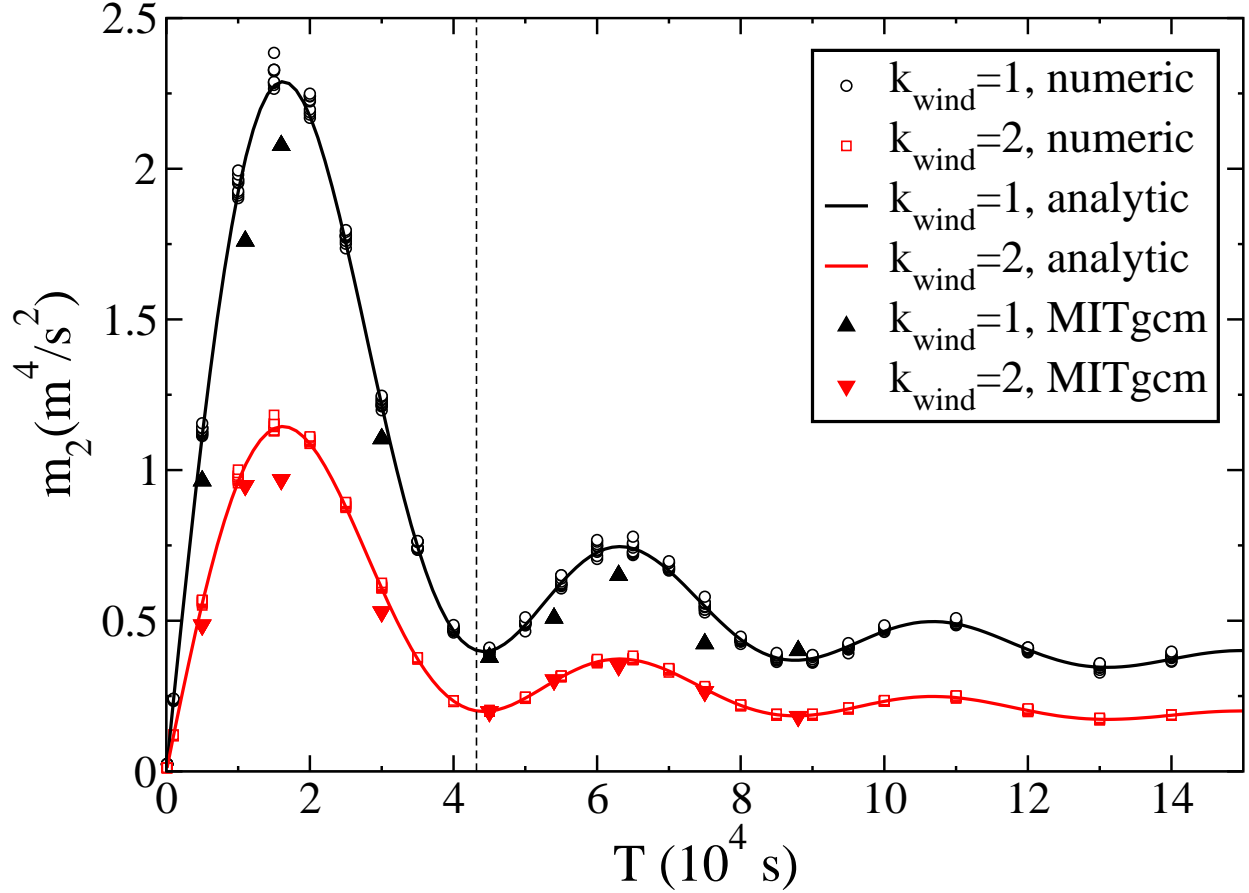}
 \caption{The second moment of the current amplitude versus the
   constant wind-stress duration, $T$.  The analytical results (solid
   lines) are compared with the numerical solution of the Ekman model
   (empty symbols) and with the MITgcm modeling
   of the currents in a simple artificial lake with open boundaries
   using a 1 $km$ resolution.  The wind-stress amplitude was drawn
   from a Weibull distribution with two different values of the
   $k_{\rm wind}$ parameter as specified in the figure.  The Coriolis
   frequency, $f\approx 1.45\times 10^{-4}s^{-1}$, was set to its
   value in the pole, and the constant wind duration associated with
   this Coriolis frequency is indicated by the vertical dashed
   line. The Rayleigh friction parameter is $r=10^{-5}s^{-1}$.}
 \label{fig:6rs}
 \end{figure}

In Fig. \ref{fig:6rs}, we show the second moment of the current
amplitude for the isotropic, step-like wind-stress model. The
analytical results (eq. \eqref{w2mT}) are compared with the numerical
solution of the Ekman model (eqs. \eqref{uvdyn}) and the MITgcm
modeling of the currents in a simple artificial lake (details of which
are provided above, using open boundaries with 1~$km$ resolution).
One can see that the dependence of the average current amplitude on
the constant wind-stress duration is non-monotonic (due to the fact
that the Coriolis effect is significant -- $|f|>r$) and that there is
an excellent agreement between all the results.  At the equator (not
shown), where $f=0$, the second moment increases monotonically to
$\langle\left|\tau\right|^2\rangle/\left(r^2\rho_0^2\right)$ as a
function of $T$.

\begin{figure}
 \centering
 \includegraphics[width=\textwidth]{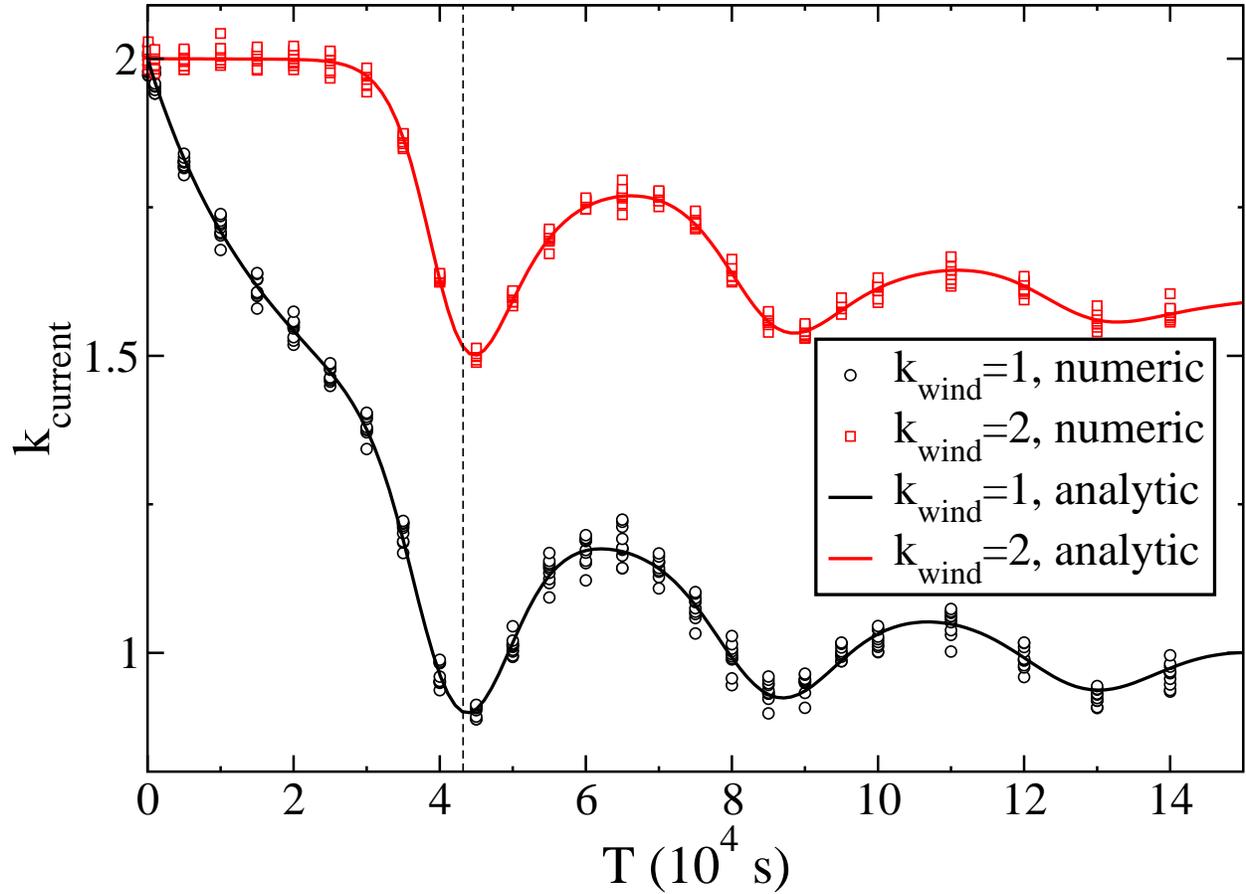}
 \caption{The Weibull $k_{\rm current}$ parameter of the current distribution versus
          the constant wind-stress duration, $T$. The analytical results (solid lines) are compared
	  with the numerical solution of the Ekman model (empty symbols).
          The wind-stress was drawn from a Weibull distribution with two
          different values of the $k_{\rm wind}$ parameter as specified in the figure.
          The Coriolis frequency, $f$, was set to its value in the pole and the 
          corresponding duration is indicated by the vertical dashed line.}
 \label{fig:7rs}
 \end{figure}

It was previously argued that, under certain conditions, the
wind-amplitude (directly related to the wind-stress) PDF is well
approximated by the Weibull distribution
\cite[][]{Monahan-2010:probability}; we thus chose, in our
demonstrations, a Weibull distribution of wind-stress.  In
Fig. \ref{fig:7rs}, we present the Weibull $k_{\rm current}$ parameter
(eq. \eqref{weibull}) of the current distribution versus the constant
wind-stress duration for two different values ($k_{\rm wind}=1$ and
$k_{\rm wind}=2$) of the wind-stress Weibull $k_{\rm wind}$ parameter
(again, we only present here the results of the isotropic
case). The analytical value of $k_{\rm current}$ was found based on the
  ratio between the fourth moment and the square of the second moment; we
  note, however, that the current's PDF is not necessarily a Weibull PDF, and
  we chose to quantify the current's PDF using the $k$ parameter of
  the Weibull distribution for presentation purposes only. The
findings of \cite{Monahan-2008:probability} are consistent with the lower
range of $k_{\rm wind}$ values used here.  One can see that, for short
correlation times, the current amplitude exhibits a Rayleigh
distribution ($k_{\rm current}=2$), independent of the wind-stress
distribution. This corresponds to Gaussian distributions of the
current components. In the other limit of long constant wind-stress
periods, $k_{\rm current}$ converges to $k_{\rm wind}$ (not shown for
$k_{\rm wind}=2$).

\begin{figure}
 \centering
 \includegraphics[width=\textwidth]{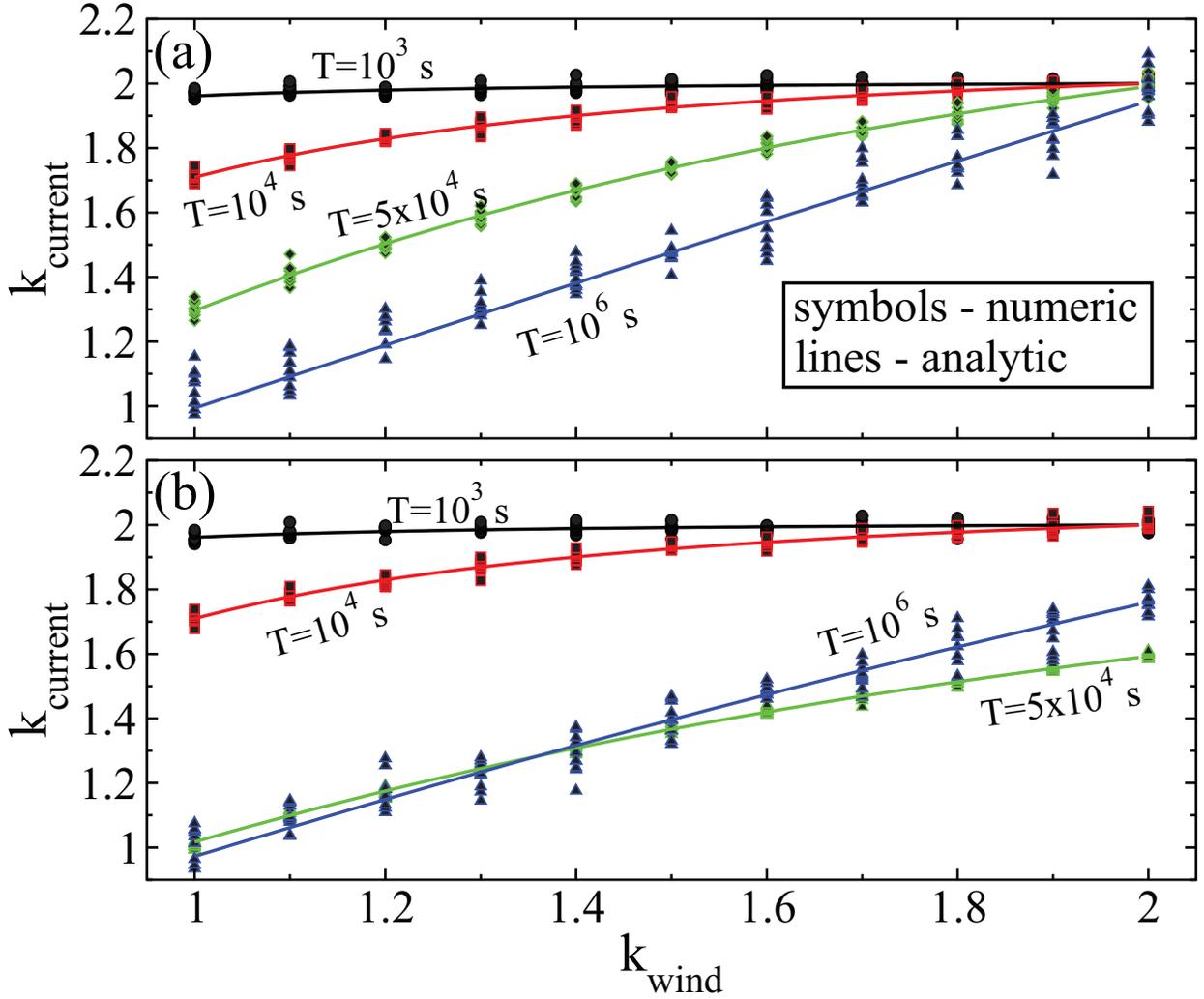}
 \caption{The Weibull $k_{\rm current}$ parameter of the current
   distribution versus the $k_{\rm wind}$ parameter of the step-like
   wind-stress distribution.  The analytical results (solid lines) are
   compared with the numerical solution of the Ekman model (symbols).
   The Coriolis parameter, $f$, in panel (a), is $0$, corresponding to
   its value at the equator.  In panel (b) , $f\approx1.45\times
   10^{-4}s^{-1}$, corresponding to its value at the pole and is
   larger than $r=10^{-5}s^{-1}$.  The different lines correspond to
   different constant wind-stress durations, $T$, as indicated in the
   figure.}
 \label{fig:8rs}
 \end{figure}

The above results were obtained for the maximal value of the Coriolis
parameter, i.e., $f$ at the pole. In Fig. \ref{fig:8rs}(a,b), we show
$k_{\rm current}$ versus $k_{\rm wind}$ for different values of the
constant wind-stress duration, $T$, and the two limiting values of the
Coriolis parameter, $f$, at the equator and at the pole. Here
again, one observes an excellent agreement between the numerical and
the analytical results for both values of $f$. The limits of short
and long temporal correlations of the wind, at which $k_{\rm
  current}=2$ and $k_{\rm current}=k_{\rm wind}$, correspondingly, are
clearly demonstrated.

\begin{figure}
 \centering
 \includegraphics[width=\textwidth]{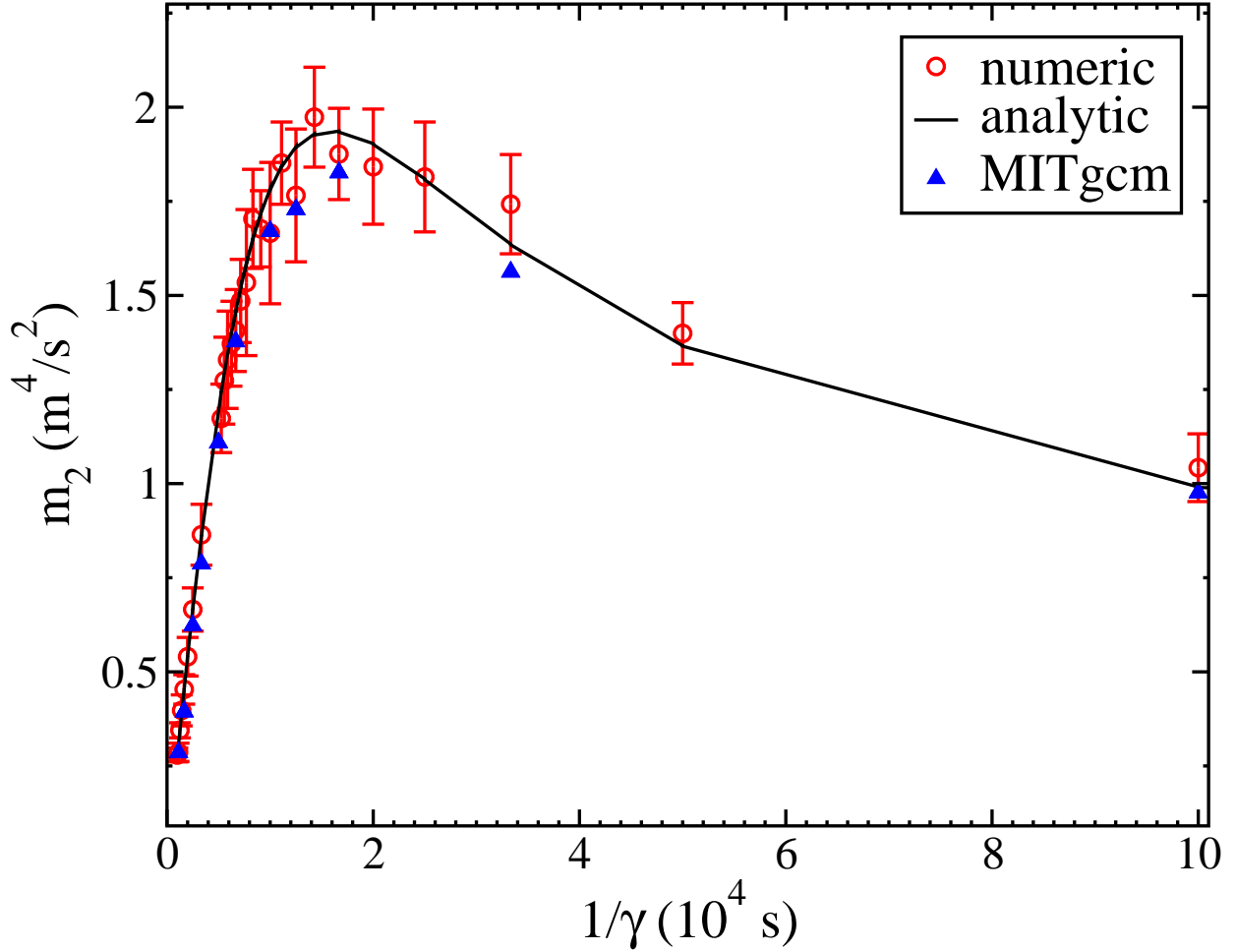}
 \caption{The second moment of the current amplitude for the case of
   wind-stress with exponentially decaying temporal correlations.  The
   analytical results (solid line) are compared with the results of
   numerical integration of Ekman equations (red circles) and with the
   currents simulated by the MITgcm model (blue triangles) in a simple
   lake with open boundaries (see the description in section
   \ref{sec:mitGCM}). The red circles show the mean of different realizations of the wind-stress in the numerical integration of the model and the bars show the standard deviation of these realizations.
   The wind-stress amplitude was drawn from a
   Weibull distribution with $k_{\rm wind}=1.5$. The Coriolis
   frequency, $f\approx 7.27\times10^{-5}s^{-1}$, corresponds to its
   value at $lat=30^\circ$ and $r=10^{-5}s^{-1}$. Note that the second
   moment is plotted versus $1/\gamma$, to allow easier comparison
   with the results of the step-like wind-stress case.  }
 \label{fig:9rs}
 \end{figure}

\begin{figure}
 \centering
 \includegraphics[width=\textwidth]{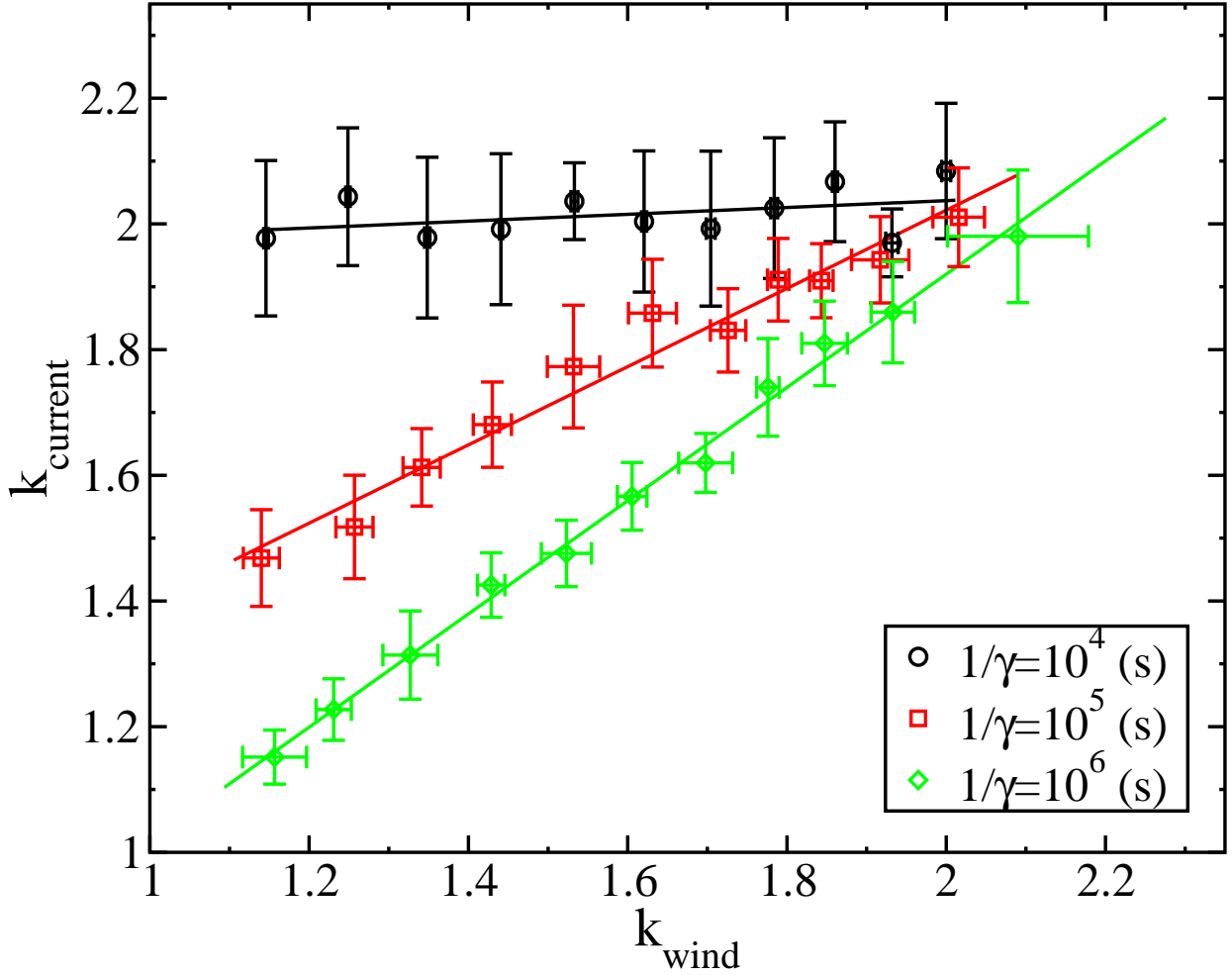}
 \caption{The Weibull $k_{\rm current}$ parameter of the current
   amplitude distribution versus the $k_{\rm wind}$ parameter of the
   wind-stress with exponentially decaying temporal correlations for
   different correlation times (corresponding to $1/\gamma$ as
   indicated in the figure). The results were obtained using different
   realizations of the wind-stress. The symbols correspond to the mean and the bars to the standard deviation of the different realizations of the wind-stress. Due to the finite size of the time series, there are also small deviations in the value of $k_{\rm wind}$, and the horizontal bars show the standard deviation of these fluctuations. The solid lines are drawn to guide
   the eye. The values of the Coriolis parameter, $f$ and the Rayleigh
   friction parameter, $r$, are the same as in Fig. \ref{fig:9rs}.}
 \label{fig:10rs}
 \end{figure}

In Figs. \ref{fig:9rs} and \ref{fig:10rs}, we present similar results
for the case of an exponentially decaying correlation function of the
wind-stress. As mentioned in Sec. \ref{sec:OUws}, the expression for
the current's second moment (eq. \eqref{2mexp}) holds for any
distribution of the wind-stress (given that the temporal
autocorrelation follows eq. \eqref{expcf}), and thus it is possible
to obtain analytically the current's second moment in the case of Weibull-distributed wind-stress whose $k$ value  does not necessarily equal 2. We generated a Weibull-
distributed time series that has exponentially decaying temporal
correlations, following the procedure described in Section
\ref{sec:weibull}.  In Fig. \ref{fig:9rs}, we present the second
moment of the current amplitude versus the correlation time (1/decay
rate) of the correlation function.  The existence of an optimal decay
rate, for which the average current amplitude is maximal, is
demonstrated for this case as well. One notable difference is the
absence of the secondary maxima points which appeared in the step-like
model. For this case, one may easily find that the maximal average
amplitude of the currents is obtained for $\gamma=|f|-r$, assuming
that $|f|\ge r$.
  
\begin{figure}
 \centering
 \includegraphics[width=\textwidth]{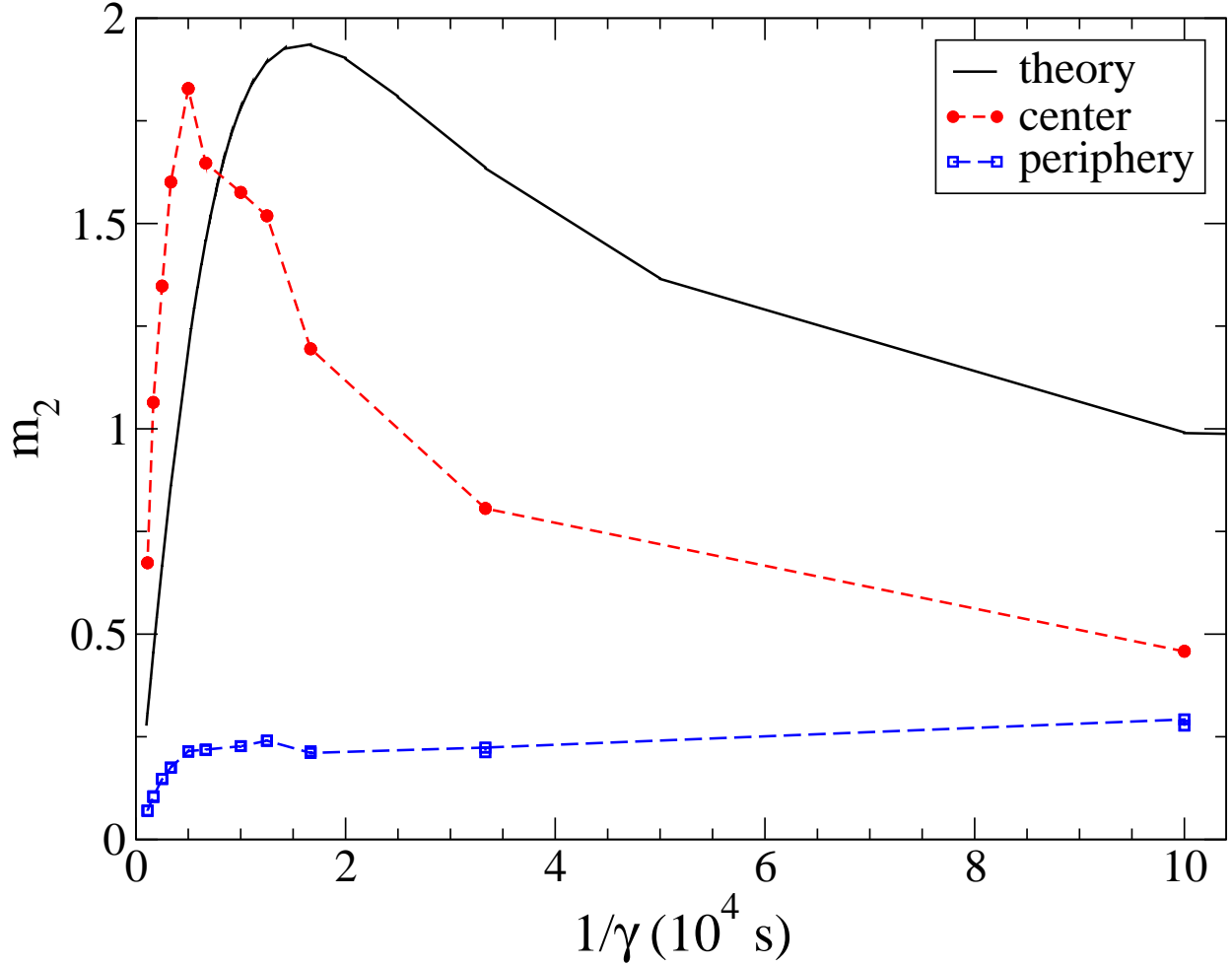}
 \caption{Comparison of the analytical results for the second moment
   of the current amplitude with the results of the detailed MITgcm
   model with closed boundaries and a coarse resolution of 10 $km$.  It
   is shown that for a wide range of the wind-stress correlation time
   (the exponentially decaying temporal correlations were considered),
   the MITgcm results agree qualitatively with the analytical results
   of the over-simplified model. On the other hand, the results in the
   periphery show a significant deviation. These results, combined
   with the fact that for open boundaries, there was a good agreement
   between the analytical and the simulation’s results, suggest that, as
   expected, the simple model fails to capture the boundary effects.}
 \label{fig:11rs}
 \end{figure}

In Fig. \ref{fig:10rs}, we present $k_{\rm current}$ versus $k_{\rm
  wind}$.  Here again, we obtain an excellent agreement between the
predicted and the numerically obtained limiting behaviors. It is
important to note that the good agreement between the MITgcm results
and the analytical results has been proven to be valid only for the
setup described above, using a fine resolution of 1 $km$ and open
boundaries. Different behavior may occur for different scenarios, such
as regions close to the boundary of the domain, spatially variable
wind, complex and steep bottom topography, and vertically and
spatially variable temperature and salinity.  Using a setup with
closed boundaries, we found that close to the boundaries of the
artificial lake, the results showed a significant deviation due to the
boundary effects that were neglected in the analytical model. The
results of the MITgcm model with closed boundary conditions and
coarser resolution of 10 $km$ are shown in Fig. \ref{fig:11rs}. All
the other figures show the MITgcm results for the case of open
boundaries. Moreover, we used a spatially uniform wind-stress in our
simulation and have not considered the more realistic case of
non-uniform wind-stress.

\section{Summary}
\label{sec:summary}

In summary, we have shown, in two idealized cases,
that the PDF of wind-driven ocean currents depends on the temporal
correlations of the wind. For short-range correlations, the current
speed approaches zero, and the PDF of its components is Gaussian. For
long-range temporal correlations of the wind, the currents' PDF is
proportional to the wind-stress PDF. The different cases considered here and the MITgcm simulations
described above suggest that the existence of a maximal current speed as a function
of the temporal correlation time is not unique to the cases studied
here, and may also be relevant in more realistic types of temporal correlations
and setups. Analysis that is based on the space-dependent model (either in the horizontal or the vertical dimensions or both) 
is a natural extension of the current study and will
allow us to compare the analytical results to altimetry-based surface
currents and to study non-local phenomena.
\begin{acknowledgments}
The research leading to these
results has received funding from the European Union Seventh
Framework Programme (FP7/2007-2013) under grant number
[293825].
\end{acknowledgments}

\newpage

\begin{thebibliography}{25}

\bibitem{Seguro-Lambert-2000:modern}
J.~V. Seguro and T.~W. Lambert,
\newblock {\em Wind Engineering and Industrial Aerodynamics} {\bf 85}, 75--84 (2000).

\bibitem{Monahan-2006:probability}
A.~H. Monahan,
\newblock {\em J. Climate} {\bf 19}, 497--520 (2006).

\bibitem{Monahan-2010:probability}
A.~H. Monahan,
\newblock {\em J. Climate} {\bf 23}(19), 5151--5162 (2010).

\bibitem{Chu-2008:probability}
P.~Chu,
\newblock {\em Geophys. Res. Lett.} {\bf 35}, L12606 (2008).

\bibitem{Gille-Smith-1998:probability}
S.~T. Gille and S.~G.~L. Smith,
\newblock {\em Phys. Rev. Lett.} {\bf 81}(23), 5249--5252 (1998).

\bibitem{Gille-Smith-2000:velocity}
S.~T. Gille and S.~G.~L. Smith,
\newblock {\em J. Phys. Oceanogr.} {\bf 30}(1), 125--136 (2000).

\bibitem{Chu-2009:statistical}
P.~Chu,
\newblock {\em IEEE J. of Selected Topics in Applied Earth Observations and
  Remote Sensing} {\bf 2}(1), 27--32 (2009).

\bibitem{Ashkenazy-Gildor-2011:probability}
Y.~Ashkenazy and H.~Gildor,
\newblock {\em J. Phys. Oceanogr.} {\bf 41}, 2295--2306 (2011).

\bibitem{Ekman-1905:influence}
V.~W. Ekman,
\newblock {\em Arch. Math. Astron. Phys.} {\bf 2}, 1--52 (1905).

\bibitem{Gill-1982:atmosphere}
A.~E. Gill,
\newblock{\em Atmosphere-Ocean Dynamics},
\newblock Academic Press, London (1982).

\bibitem{Cushman-Roisin-1994:introduction}
B.~Cushman-Roisin,
\newblock{\em Introduction to Geophysical Fluid Dynamics},
\newblock Prentice Hall, 1st edition (1994).

\bibitem{Airy-1845:tides}
G.~B. Airy,
\newblock {On Tides and Waves},
\newblock {\em Encyclopedia Metropolitana} {\bf 5}, 241-396 (1845).

\bibitem{Pollard-Millard-1970:comparison}
R.~T. Pollard and R.~C. Millard,
\newblock {\em Deep Sea Res.} {\bf 17}, 813--821 (1970).

\bibitem{Kundu-1976:analysis}
P.~K. Kundu,
\newblock {\em J. Phys. Oceanogr.} {\bf 6}, 879--893 (1976).

\bibitem{Kase-Olbers-1979:wind}
R.~H. Kase and D.~J. Olbers,
\newblock {\em Deep Sea Res.} {\bf 26}, 191--216 (1979).

\bibitem{Simons-1980:circulation}
T.~J. Simons,
\newblock {\em Can. Bull. Fish. Aquat. Sci.} {\bf 203}, 146 (1980).

\bibitem{vanKampen-1981:stochastic}
N.~G. van Kampen,
\newblock{\em Stochastic Processes in Physics and Chemistry},
\newblock North-Holland (1981).

\bibitem{Monahan-2007:empirical}
A.~H. Monahan,
\newblock {\em J. Climate} {\bf 20}, 5798--5814 (2007).

\bibitem{Monahan-2012:temporal}
A.~H. Monahan,
\newblock {\em J. Climate} {\bf 25}, 6684--6700 (2012).

\bibitem{Gonella-1971:local}
J.~Gonella,
\newblock {\em Deep Sea Res.} {\bf 18}, 775--788 (1971).

\bibitem{MITgcm-manual-2010:mitgcm1}
MITgcm group, MITgcm User Manual, Online documentation, MIT/EAPS (2010).
\newblock {\em http://mitgcm.org/public/r2 manual/latest/online documents/manual.html}

\bibitem{SchreiberAndSchmidt:1996}
T.~Schreiber and A. Schmitz,
\newblock {\em Phys. Rev. Lett.} {\bf 77}, 635 (1996).

\bibitem{SchreiberAndSchmidt:2000}
T.~Schreiber and A. Schmitz,
\newblock {\em Physica D} {\bf 142}, 346 (2000).

\bibitem{KantzAndSchreiber:2004}
H.~Kantz and T.~Schreiber,
\newblock {\em Nonlinear Time Series Analysis},
\newblock Cambridge University Press, Cambridge (2004).

\bibitem{McWilliams-Huckle-2005:ekman}
J.~C. McWilliams and E.~Huckle,
\newblock {\em J. Phys. Oceanogr.} {\bf 36}, 1646--1659 (2005).

\bibitem{Monahan-2008:probability}
A.~H. Monahan,
\newblock {\em Geophys. Res. Lett.} {\bf 35}, L05704 (2008).

\bibitem{Kalnay-Kanamitsu-Kistler-et-al-1996:ncep}
E.~Kalnay et~al,
\newblock {\em Bulletin of the American Meteorological Society}
  {\bf 77}(3), 437--471 (1996).

\end{thebibliography}
\end{document}